\documentstyle[12pt]{article}
\newcommand{\Ir}{Z\!\!\!Z}

\newcommand{\Ibb}[1]{ {\rm I\ifmmode\mkern
            -3.6mu\else\kern -.2em\fi#1}}
\newcommand{\ibb}[1]{\leavevmode\hbox{\kern.3em\vrule
     height 1.2ex depth -.3ex width .2pt\kern-.3em\rm#1}}
\newcommand{\Cx}{{\ibb C}}

\newcommand{\Rl}{{\Ibb R}}
\newcommand{\ins}{\leavevmode
    \vbox{\kern.2em \hrule width1.2ex height0.1ex}
    \hbox{\vrule width0.1ex height1.2ex depth0.ex \kern.1em} }
\newcommand{\be}{\begin{eqnarray}}
\newcommand{\ee}{\end{eqnarray}}
\newcommand{\bez}{\begin{eqnarray*}}
\newcommand{\eez}{\end{eqnarray*}}
\renewcommand{\O}{\Omega}
\newcommand{\A}{{\cal A}}
\newcommand{\X}{{\cal X}}

\renewcommand{\d}{\mbox{d}}
\newcommand{\D}{\mbox{D}}

\newcommand{\bu}{\bullet}
\newcommand{\pa}{\partial}
\newcommand{\na}{\nabla}
\newcommand{\we}{\wedge}
\newcommand{\hp}{\hat{\partial}}
\newcommand{\tp}{\tilde{\partial}}
\newcommand{\<}{\langle}
\renewcommand{\~}[1]{\tilde{#1}}
\renewcommand{\>}{\rangle}

\newcommand{\lb}{\lbrack}
\newcommand{\rb}{\rbrack}
\begin{document}
\renewcommand{\theequation} {\arabic{section}.\arabic{equation}}
\centerline{\huge \bf Non-commutative Geometry and}
\vskip.5cm
\centerline{\huge \bf Kinetic Theory of Open Systems}
\vskip2.cm
\begin{center}
\begin{minipage}{13cm}
{\bf A. Dimakis}$^\dagger$ and {\bf C. Tzanakis}$^\ddagger$
\vskip .3cm
$^\dagger$ Department of Mathematics, University of Crete,
GR-71409 Iraklion   \\
$^\ddagger$ Department of Education, University of Crete,
GR-74100 Rethymnon
\end{minipage}
\end{center}
\vskip2.cm

\begin{abstract}
\noindent
The basic mathematical assumptions for autonomous linear kinetic equations
for a classical system are formulated, leading to the conclusion that if
they are differential equations on its phase space $M$, they are at most
of the 2nd order. For open systems interacting with a bath at canonical
equilibrium they have a particular form of an equation of a generalized
Fokker-Planck type. We show that it is possible to obtain them as Liouville
equations of Hamiltonian dynamics on $M$ with a particular non-commutative
differential structure, provided certain geometric in character, conditions
are fulfilled. To this end, symplectic geometry on $M$ is developped in this
context, and an outline of the required tensor analysis and differential
geometry is given. Certain questions for the possible mathematical
interpretation of this structure are also discussed.
\end{abstract}

\section{Introduction}
\setcounter{equation}{0}

The understanding and description of irreversible evolution of macroscopic
systems, presumably towards equilibrium states is the aim of kinetic theory.
This is done by formulating appropriate kinetic equations giving the time
evolution of the state of the system which loosely speaking, is assumed to be
some probability measure on the space of variables describing the system,
which henceforth will be called its phase-space. Observables are assumed to
be well-behaved phase-space functions and experimental results refer to
expectation values obtained via a bilinear (or sesquilinear) form on the
cartesian product of the spaces of states and observables. Thus at the
very heart of kinetic theory a probabilistic view point is rooted, the
precise interpretation of which is often a matter of debate.

Theoretically speaking, kinetic equations are derived by following two
different procedures: \begin{enumerate}
\item[(i)] stochastic arguments, based on some presumably plausible
assumptions on the behaviour of a large number of microscopic events
characterising the system.
\item[(ii)] application of more or less systematic approximation schemes
on the exact microscopic dynamics of the system under consideration.
\end{enumerate}

Typical examples for (i) are the Fokker-Planck and Kramers equations
(e.g.\ \cite{1} ch.VIII). The first refers to the probability
distribution function (p.d.f.) of the velocity of a heavy Brownian particle
suspended in an equilibrium medium of light particles, the second one, to
the p.d.f.\ of its phase-space position under the presence of an external
field. The basic assumptions for their derivation are that the microscopic
dynamics of the Brownian particle are governed by Langevin's equation and
that its p.d.f.\ is a Markovian diffusion process (see e.g.\ \cite{2}
ch.II, \cite{2a}).
This procedure usually leads to {\em model} equations
that are easier to study than the {\em assumed} more fundamental equations
following from exact dynamics by (ii). Mathematically speaking, this
procedure essentially {\em replaces exact dynamics by stochastic differential
equations} which in turn imply evolution equations for the p.d.f.\ of the
system (e.g.\ \cite{3} ch.4, \cite{4} ch.9).
Though interesting and important from both the mathematical and physical
point of view, it seems that one has to know somehow the kinetic equation
he wants to derive and modify accordingly the corresponding dynamics.
Moreover the modification is not easily interpreted physically
(see the discussion following (\ref{3.1}) below). That is, it seems that
there is no general prescription of how this modification has to be carried
out. In this paper, starting from mathematically totally different concepts
we do arrive at results that may be interpreted stochastically, at the same
time giving hints on the nature of such a general prescription
(cf.\ the discussion following (\ref{3.1}) below).

Typical examples for (ii) are the Boltzmann equation for dilute classical
gases (or its quantum weak-coupling analogue, Pauli's equation) or the
Landau and Balescu-Lenard equations for neutral plasmas, as they are derived
from Liouville's (or von Neumann's) equations by using either iteration
schemes and projection operator methods, or by using its equivalent,
the BBGKY hierarchy of equations truncated on the basis of physical
considerations.\footnote{See e.g.\ \cite{5} \S\S30, 41, \cite{6} ch.20,
\cite{7} ch.IX, \cite{8} \S2.4, \cite{9} \S\S4.3, 4.6, \cite{10}
{\S}VI.}

Although many specific equations can be derived by mathematically
satisfactory (or even rigorous) methods in some particular limit of
appropriate parameters of the system (for a survey see \cite{32}), it is
true that {\em any approximation scheme, leading to satisfactory kinetic
equations for particular classes of systems, runs into trouble as long as
one tries to extend it to other systems and/or higher order of
approximation} (\cite{11}, \cite{12} \S5). For instance the linearized
Landau equation follows for spatially homogeneous plasmas as a 2nd order
approximation in the plasma parameter to the Liouville equation (plus
some additional assumptions which need not be discussed here). Any effort
to find its generalization either in higher order of approximation and/or
for inhomogeneous systems, runs into difficulties (e.g.\ equations
violating the positivity of p.d.f or having no $H$-theorem, are obtained
--- see e.g.\ \cite{13} in connection with \cite{12} \S5), or involves
highly ad hoc steps which are sometimes hidden in the calculations
(see e.g.\ \cite{14} in connection with \S2 below).

Although one may be tempted to accept that there is no reason to expect that
equations in an approximate theory share all properties of the corresponding
equations in the exact theory, in kinetic theory  things are more complicated
since the exact theory in this case is simply (classical or quantum) dynamics;
however as it stands, microscopic dynamics of a system with a very large
number of degrees of freedom is useless as a macroscopic theory since it does
not incorporate irreversible evolution and its relation to a theory dealing
with macroscopically defined quantities is remote, or at least not
straightforward.  Moreover from what has been said above, the situation is
even worse, since it often happens that equations obtained at a lower level
of approximation (e.g.\ with respect to expansion in some parameter)
exhibit the correct properties, which however disappear in any higher
level of approximation!\footnote{See e.g.\ equations following from expansion
of
Liouville's equation or the so-called generalized master equation \cite{9}
\S2.3, \cite{11} \S2, \cite{12} \S5, \cite{26} in connection with the last
paragraph of the next section. See also \cite{33} for expansions of the
Boltzmann equation.}

In our opinion this is an inevitable consequence of the philosophy underlying
kinetic theory, namely that {\em irreversible evolution} is simply an
approximation to the exact (classical or quantum) {\em reversible dynamics}.
To put it differently, the {\em aim} of kinetic theory (description of
macroscopic irreversible evolution) seems not to be {\em compatible} with
its {\em assumed fundamental laws} (reversible microscopic dynamics), at
least in a straightforward manner. Therefore it seems necessary that a
fundamentally different approach to kinetic theory may not be worthless!

\section{Mathematical assumptions underlying kinetic theory
and their implications.}
\setcounter{equation}{0}

As already mentioned, at the very root of kinetic theory lies a
probabilistic interpretation. Using this as a motivation and in order to
clarify the difficulties noticed in the previous paragraph, we here describe
the basic assumptions for a kinetic equation to be in principle, acceptable,
and conditions under which an explicit general form can be obtained.
We restrict the discussion to {\em linear, autonomous} kinetic equations,
which include kinetic equations for the important class of {\em open systems}.
To be more specific, we consider {\em classical} systems, though similar
results are known  for quantum systems as well (\cite{18}, \cite{15}
Theorem 4.2).

We assume that \begin{enumerate}
\item[(i)] The phase-space $M$ is a locally compact, Hausdorff topological
space (e.g. the phase-space of a Hamiltonian system).
\item[(ii)] Observables $A$ are in $C(M,\Cx)$ the space of complex-valued
continuous functions on $M$, having a finite limit at infinity.
\item[(iii)] States $\ell$ are positive linear functionals on the
observables, their values $\ell(A)$ giving expectations. Since positivity
of $\ell$, implies its boundedness in the supremum norm (e.g. \cite{20}
p.106--107), states belong to $C^\ast(M,\Cx)$, the Banach dual with
respect to this norm, which is the space of (regular) complex  Borel
measures on $M$.
\item[(iv)] Kinetic equations for the observables have well-posed initial
value problem, i.e.\ uniqueness and continuous dependence of solutions on
the initial data hold. Moreover, expectation values are continuous in time.
These imply that solutions of a kinetic equation define strongly
continuous semigroups of linear operators on the space of observables and
that the corresponding adjoint equation defines such a semigroup on the
state space.\footnote{Strictly speaking, strong continuity of the latter
holds on a smaller subspace, which however uniquelly defines the adjoint
semigroup (\cite{21} Theorem 1.36).}
\item[(v)] The adjoint semigroup conserves positivity and normalization
of the states, i.e.\ initial probability measures retain their character
for all positive times.
\end{enumerate}

These plausible assumptions imply that the solutions of a kinetic equation
for the observables define a Markov semigroup, i.e.\ a strongly continuous,
positivity and normalization preserving one-parameter semigroup of operators
on $C(M,\Cx)$ \cite{26}. The terminology stems from the fact that such
semigroups are in one to one correspondence with (time-homogeneous)
stochastically continuous Markov processes described by a transition
probability distribution $p(t,x,E)$, which for each $t,x$ is a regural
probability Borel measure on $M$ (see e.g.\ \cite{22} p.399, \cite{23}
Theorem 2.1, for outline of a proof). If we further {\em assume} that
$M$ is an $n$-dimensional differential manifold, the generator of the
semigroup is defined on $C^2$-functions and that a Lindenberg's type
condition holds,
\bez
   \lim_{t\to 0^+}{p(t,x,E)\over t}=\chi_E(x)\qquad\qquad
   \mbox{uniformly in $x$,}
\eez
where $\chi_E$ is the characteristic function of $E$, then it can be proved
that the kinetic equation for observables (i.e.\ essentially the generator
of the corresponding semigroup) has the form\footnote{See \cite{24}
Theorem 5.3 for outline of a proof, and \cite{25} Theorem XIII.53 for a
partial generalization --- the L\'{e}vy-Khinchine formula.}
\be
 {\pa A\over\pa t}=\alpha^{ij}(x)\pa_i\pa_j A+ a^i(x)\pa_i A\;,\label{2.1}
\ee
where $\alpha^{ij}$, $a^i$ are continuous, $\alpha^{ij}$ is a non-negative
definite matrix function and the summation convention has been used, as
it will be done in the sequel.\footnote{For detailed proofs and more precise
formulation of the various conditions see \cite{26}.}
Actually $\alpha^{ij}, a^i$ are related to the diffusion and drift
coefficients associated with the corresponding Markov process, given by
the first two moments of $p$.

In many approaches in kinetic theory which lead to satisfactory kinetic
equations of the form (\ref{2.1}), the following condition holds
\be  a^i=b^i+X_H^i\;, \label{2.2} \ee
where $X_H^i$ is a {\em Hamiltonian} vector field. Moreover for {\em open}
(classical or quantum) systems with Hamiltonian $H$, in interaction with
``baths'' in canonical equilibrium, the corresponding Hamiltonian function
is an integral of the unperturbed motion of the open system, and therefore
has in general the form $-(H+F)$ with
\be  \{ F,H \}=0\;,\label{2.3} \ee
$\{\,,\}$ being the Poisson bracket, or the operator commutator, and the minus
sign gives the correct Hamiltonian equation when the system does not
interact with the bath.

Moreover
\be \pa_j\alpha^{ij}-b^i=\beta\alpha^{ij}\pa_j H\;, \label{2.4}\ee
where $\beta$ is proportional to the inverse temperature of the
bath.\footnote{See e.g.\ \cite{12} Proposition 4.1, \cite{27} eq.(2.19),
\cite{5} p.190, \cite{17} eqs(III.26) and (III.16) together with (III.19),
\cite{31} eq.(5.8).} These conditions will be used in section 4.

Substitution of (\ref{2.2})--(\ref{2.4}) in (\ref{2.1}) gives
\be
   {\pa A\over \pa t}=-\{H+F,A\}+\pa_i(\alpha^{ij}\pa_j A)-
   \beta\alpha^{ij}\pa_j H\pa_i A\;.\label{2.5}
\ee
The essential conclusion drawn from the above discussion is (in a somewhat
nonrigorous language) that {\em linear autonomous kinetic equations for
classical systems, which are differential equations are necessarily at
most of the 2nd order with nonnegative-definite leading coefficient and
vanishing zeroth order term (cf.\ eq.(\ref{2.1}))}\,.\footnote{A corresponding
result is also known for quantum systems \cite{18}, \cite{16}, \cite{17}.
The special case of the Kramers-Moyal expansion of the linearized
Boltzmann equation has been considered in \cite{33}.}

Since kinetic equations following from microscopic dynamics (see \S1)
are often differential equations, the above result, severely restricts their
form. In particular all methods based on power expansions of the solution of
Liouville's equation with respect to some appropriate parameter, usually
lead to unacceptable results since each approximation step increases
the order of the differential operator by one.\footnote{See e.g.\ \cite{9}
\S2.4 eq.(2.199), \cite{1} p.215, {\S}IX.6 particularly p.280.}
In the rest of this paper, we will present a different point of view,
motivated by the discussion in the next section.

\section{Non-commutative geometry and stochastic calculus.}
\setcounter{equation}{0}

The usual approach to kinetic theory (method (ii) in Section 1) follows the
scheme
\bez
\begin{array}{ccc}
\mbox{Microscopic (Hamiltonian dynamics)} & & \\
  +  &  \Longrightarrow & \mbox{Kinetic equations} \\
\mbox{Some systematic approximation scheme} & &
\end{array}
\eez
On the other hand, method (i) in Section 1, involves no systematic
approximation scheme but a modification of microscopic dynamics which
become now stochastic differential equations. In our opinion, the weak
point in this case is that there is no general method of how to choose the
microscopic stochastic differential equations.  That is, it seems necessary
that one somehow knows the kinetic equation he wants to derive and then
writes down the corresponding stochastic equation. However it is not always
clear how to interpret the latter.

For instance it would be desirable to be able to derive by method (i) kinetic
equations obtained from microscopic Hamiltonian dynamics. However, the
latter in general involve operators with derivatives in the $q$-coordinates
(see e.g.\ \cite{11}, \cite{12}, \cite{27}, \cite{31}). And it is a standard
fact that stochastic differentil equations in phase-space involving Wiener
processes $\vec{X}_t$ of the form
\be
 \left(\begin{array}{c}	d\vec{q} \\
                        d\vec{p} \end{array}\right) =
 \left(\begin{array}{c}
\vec{A}_q(\vec{q},\vec{p})\, dt+{\bf F}_q(\vec{q},\vec{p})\cdot d\vec{X}^q_t\\
\vec{A}_p(\vec{q},\vec{p})\, dt+{\bf F}_p(\vec{q},\vec{p})\cdot d\vec{X}^p_t
                                \end{array}\right)\;, \label{3.1}
\ee
imply such derivatives in the corresponding kinetic equation if ${\bf F}_q
\neq 0$ (see e.g.\ \cite{3} \S4.3.3). However in many cases, one would
like to interpret $d\vec{q}/dt$ simply as a velocity, all dynamics being
incorporated in $d\vec{p}/dt$ (e.g. the velocity of a Brownian particle),
in which case the above stochastic equation (\ref{3.1}) is not easily
interpreted physically.

In the next sections we outline {\em another approach} in which {\em the
Hamiltonian character of microscopic dynamics is retained, but instead of
approximation schemes, we make the fundamental assumption that observables
are now defined on a manifold with noncommutative geometrical structure}\,.
As it will be explained, this may be interpreted as a stochastic dynamical
structure, though it is not known if this is necessary.
Nevertheless, the formulation developped in the next section is closely
related to what may be called a differential geometric approach to
stochastic calculus\footnote{See e.g.\ \cite{36}, \cite{37} particularly
ch.VI, compare also with the approach in \cite{38}.} though this relation
will be further explored in another paper.

The {\em mathematical} motivation for introducing noncommutative geometrical
structure, stems from the fact that in the one-dimensional version of
(\ref{3.1}), $X_t$ being a Wiener process (i.e.\ Langevin's equation)
the usual associative product between differential forms spanned by $dt$,
$dX_t$ and functions of $t$, $X_t$, implies that stochastic differentiation
does not obey Leibniz's rule for the product of two functions, due to the
appearance of 2nd derivatives in It\^o's formula for the differential of
such functions (\cite{28} \S2, particularly eq.(2.4)).However it is possible
to define  a modified {\em noncommutative} but still associative product
between functions and differential forms such that Leibniz's rule holds
(\cite{28} section 3). This product induces a noncommutative differential
calculus on the ordinary algebra of functions of $t$, $X_t$ (\cite{29}
section 2) via the basic commutation relations between functions and 1-forms
\be \begin{array}{r}
   \lb \d t,t \rb =\lb \d t,X_t\rb =0\;, \\
   \lb\d X_t,X_t \rb =2\gamma\,\d t\;, \end{array}\label{3.2}
\ee
$\gamma$ being the diffusion constant appearing in the usual Fokker-Planck
equation obtained from Langevin's equation (\cite{28} eq.(3.16)).
Eq.(\ref{3.2}) is readily generalized to a multi-dimensional Wiener process:
\be \begin{array}{r}
   \lb\d t,t \rb =\lb\d t,X^i_t\rb =0\;,  \\
   \lb\d X^i_t,X^j_t \rb =b^{ij}\,\d t\;,\end{array} \label{3.3}
\ee
where $b^{ij}$ is a symmetric bilinear form on the space of
1-forms.\footnote{This is implied by the fact that since $\d$ satisfies
Leibniz's rule, $\lb\d X^i_t,X^j_t \rb = \lb\d X^j_t,X^i_t \rb$ which
in fact shows that this commutator depends only on $\d X^i_t$, $\d X^j_t$
--- see \cite{29} \S3.} Therefore (\ref{3.3}) can be rewritten more generally
as
\be \lb \d f, g\rb = b(\d f,\d g)\,\d t\;, \label{3.4}
\ee
for functions $f,g$ of $\vec{X}_t,t$ and a symmetric, bilinear form with
components $b^{ij}$ in the ``coordinates'' $X^i_t$ and such that $\d t$ lies
in its kernel, i.e.\ $b^{tt}=b^{it}=0$.

In view of the above discussion, we consider in the next section a
$(2n+1)$-dimensional manifold $M\times\Rl$ and a differential calculus
on a subalgebra $\A$ of the algebra of complex-valued functions on
$M\times\Rl$ satisfying (\ref{3.4}), and outline {\em the formulation of
extended Hamiltonian dynamics as symplectic geometry on $M\times\Rl$}.
The basic result is that {\em Liouville's equation for observables turns out
to be of the form (\ref{2.1}) with conditions (\ref{2.2}), (\ref{2.4})
having a simple geometrical meaning. Therefore it may be interpreted as a
kinetic equation on the space of observables corresponding to a classical open
system with phase-space $M$}.

The construction is coordinate-independent, and presupposes the definition of
such fundamental concepts as vector fields, linear connections, symplectic
structure and antisymmetric wedge product of forms on the differential
calculus defined on $\A$ by (\ref{3.3}), in close analogy with the
corresponding concepts of the ordinary differential geometry.\footnote{For
details on the systematic definition and presentation of general results
in noncommutative geometry on a commutative algebra, see \cite{29},
particularly \S\S2,3.} The derivation is {\em formal} in the sense that no
systematic study of the representation theory of (\ref{3.3}) is made. Its
already mentioned relation with stochastic calculus is a possibility,
but it is not clear if others exist. A preliminary discussion of this problem
is given in section 5.

To make the presentation as transparent as possible, detailed calculations
will be given in a subsequent paper in which tensor analysis for the
corresponding noncommutative differential calculus is developped
systematically. In fact this work constitutes only a first step towards
a systematic formulation of kinetic theory as Hamiltonian (symplectic)
dynamics in phase-space equipped with a noncommutative geometrical
structure.

\section{Noncommutative symplectic geometry}
\setcounter{equation}{0}

Let $M$ be a $2n$-dimensional manifold and $\A$ the algebra of smooth
functions on $M\times\Rl$. The coordinate on $\Rl$ will be denoted by
$t$. Let $\~{\O}$ be the universal differential envelope of $\A$, i.e.\
$\~{\O}$ is a $\Ir$-graded algebra $\~{\O}=\bigoplus_{r\in\Ir}\~{\O}^r$
with $\~{\O}^r=\{0\}$ for $r<0$ and $\~{\O}^0=\A$. Then there exists a linear
mapping $\~{d}:\~{\O}\to\~{\O}$ of grade 1, which satisfies
\begin{enumerate}
\item[(i)] $\~{d} 1=0$, where 1 is the constant function with value 1,
\item[(ii)] $\~{d}$ satisfies the graded Leibniz rule i.e.\
 $\~{d}(\psi\psi')=(\~{d}\psi)\psi'+(-1)^r\psi(\~{d}\psi')$, for
 $\psi\in\~{\O}^r$,
\item[(iii)] $\~{d}^2=0$ on all of $\~{\O}$ and
\item[(iv)] $\A$ and $\~{d}\A$  generate $\~{\O}$.
\end{enumerate}

The universal differential envelope $(\~{\O},\~{d})$ of $\A$ can be
realized as follows (see \cite{39}): think of $\phi\in\~{\O}^r$ as a function
on
$(M\times\Rl)^{r+1}$, where for $f\in\~{\O}^0$ and $a\in M\times\Rl$
$f(a)$ is the value of $f$ as an element of $\A$ on $a$ and
for $a_0,\ldots,a_{r+1}\in M\times\Rl$ and $\phi\in\~{\O}^r$ we set
\be
 (\~{d}\phi)(a_0,\ldots,a_{r+1}):=\sum_{k=0}^{r+1}
  (-1)^k\phi(a_0,\ldots,a_{k-1},a_{k+1},\ldots,a_{r+1})\;.\nonumber
\ee
Furthermore for $\phi\in\~{\O}^r$, $\psi\in\~{\O}^s$ and $a_0,\ldots
a_{r+s}\in M\times\Rl$ we set
\be
 (\phi\psi)(a_0,\ldots,a_{r+s}):=[\phi(a_0,\ldots,a_r)]
    [\psi(a_r,\ldots,a_{r+s})]\;, \nonumber
\ee
for any non negative integers $r,s$. According to these rules
\be
(f\~{d} g\,h)(a,b)=f(a)[g(b)-g(a)] h(b)\;,\label{4.1}
\ee
and hence $\~{d} f\,g\neq g\~{d} f$. On the $\A$-bimodule of 1-forms
$\~{\O}^1$ we define a new product $\~{\bu}:\~{\O}^1\times\~{\O}^1
\to\~{\O}^1$ as follows: for $\alpha,\beta\in\~{\O}^1$ and
$a,b\in M\times\Rl$ we set
\bez
(\alpha\~{\bu}\beta)(a,b):=\alpha(a,b)\,\beta(a,b)\;.
\eez
Note that
\bez
(f_1\alpha f_2)\~{\bu}(g_1\beta g_2)=f_1g_1(\alpha\~{\bu}\beta)f_2g_2\;,
\eez
and $[\~{d} f,g]=\~{d} f\~{\bu}\~{d} g$.
The universality of $(\~{\O},\~{\d})$ is expressed by the property that, if
$(\O,\d)$ is any differential algebra on $\A$, then there is a
graded-algebra homomorphism $\pi:\~{\O}\to\O$ of grade 0 such that
$\pi|\~{\O}^0=\mbox{id}_\A$ and $\d\circ\pi=\pi\circ\~{\d}$
(cf.\ \cite{29} \S3.1).

Let $\~{b}:\~{\O}^1\times\~{\O}^1\to\A$ be a symmetric left-right
$\A$-bilinear form i.e.\ $\~{b}(f_1\alpha f_2,g_1\beta g_2)=
f_1g_1\,\~{b}(\alpha,\beta)\,f_2g_2$  and assume that $\~{d}t$ lies in
the kernel of $\~{b}$.
Let also ${\cal I}$ denote the differential ideal of $\~{\O}$ generated by
$\alpha\~{\bu}\beta-\~{\d}t\,\~{b}(\alpha,\beta)$, then we set
$\O:=\~{\O}/{\cal I}$ and $\pi:\~{\O}\to\O$ for the canonical projection.
Since ${\cal I}$ is differential the operator $\d:\O\to\O$ given by
$\d=\pi\circ\~{\d}$ is well defined, and because $\~{\d}t$
lies in the kernel of $\~{b}$, a symmetric left-right $\A$-bilinear form
$b:\O\times\O\to\A$ is uniquelly defined by $b\circ(\pi\times\pi)=\~{b}$.

Now set
\be \d f\bu\d g:=[\d f,g] \;, \label{4.2} \ee
and extend by left-right $\A$-bilinearity, then it is easy to see that
$\pi(\~{\alpha}\~{\bu}\~{\beta})=\alpha\bu\beta$, where
$\alpha=\pi(\~{\alpha})$ and similarly for $\beta$. Obviously we have
\be \alpha\bu\beta=\d t\,b(\alpha,\beta)\;. \label{4.3}\ee
This is but a special case of the general procedure used to relate
$(\~{\O},\~{\d})$ to any other differential calculus $(\O,\d)$ via
an $\A$-bimodule homomorphism $\pi$, with ${\cal I}=\ker\,\pi$
and induce $\bu$ on $\O^1$ by $\~{\bu}$ on $\~{\O}^1$
(cf.\ \cite{29} \S 3.2).
Let $\xi^i,\,i=1,\ldots,2n$ be local coordinates on $M$ then the elements
of $\A$ can be written locally as functions of $t,\,\xi^i,\,i=1,\ldots,2n$
(see also section 5). If we set $b^{ij}:=b(\d\xi^i,\d\xi^j)$ we find the
commutation relations --- note that $b(\d t,\d\xi^i)=b(\d\xi^i,\d t)=0$ ---
\be
    \lb \d\xi^i,\xi^j \rb & = & \d t b^{ij}\;, \label{4.4}\\
  \lb \d t,t \rb = \lb \d t,  \xi^i \rb = \lb \d\xi^i, t \rb  & = & 0\;.
           \label{4.5}
\ee
These are special cases of
\be    [\d f, g] = \d t \, b(\d f,\d g)\;.  \label{4.6}\ee
By (\ref{4.2}), (\ref{4.5}) we get for any $f,g,h\in\A$ that
\be \d f\bu\d g\bu\d h=0\;. \label{4.7}\ee
Applying $\d$ on a product of two functions and using (\ref{4.6}) we obtain
\be  \d(fg)=(\d f)g + (\d g)f - \d t\,b(\d f,\d g)\;. \label{4.8} \ee
Considering $\O^1$ as a right $\A$ module the dual module $\X$ is a
left $\A$ module. We write $\langle X,\alpha\rangle$ for the duality
contraction. If we define
\be  Xf:=\langle X,\d f\rangle \;, \label{4.9}\ee
then we obtain from (\ref{4.8})
\be X(fg)=g(Xf)+f(Xg)- b(\d f,\d g) (Xt)\;. \label{4.10}\ee
The elements  of $\X$ will be called {\em vector fields}. It can be proved
that as a left $\A$ module $\X$	is free with basis
given by $\hp_t,\,\pa_1,\ldots,\pa_{2n}$, where
\be\begin{array}{c}
   \pa_i:=\displaystyle{\pa\over\pa\xi^i}\;,\\
   \hp_t:=\pa_t-{1\over2}b^{ij}\pa_i\pa_j\;.
\end{array}\label{4.11}
\ee
Thus for $X\in\X$ we have
\be X=\hat{X}^t\,\hp_t + X^i\,\pa_i\;, \label{4.12}\ee
with $X^i:=(X\xi^i)$, $\hat{X}^t:=(Xt)$. More generally it can be proved
that for any differential calculus on a differential manifold, satisfying
(\ref{4.7}) vector fields, i.e.\ elements of $\X$, are second order
differential operators without constant term, like (\ref{4.12}). Thus the
name 2nd order calculus is justified in this case (cf.\ section 5).
As a further consequence $\O^1$ is
free with dual basis $\d t,\,\d\xi^i,\,i=1,\ldots,2n$, and hence
\be \d f=\d t\,\hp_t f+\d\xi^i\,\pa_i f\;.\label{4.13} \ee
Using the bilinear form $b$ we define a linear mapping from $\O^1$ to $\X$,
$\alpha\mapsto\alpha^b$ by
\bez \langle \alpha^b,\beta\rangle:=b(\alpha,\beta)\;. \eez
Note that $(\d t)^b=0$ and $(\d\xi^i)^b=b^{ij}\pa_j$.

Relations for forms of higher grade are obtained by applying $\d$ on
equations (\ref{4.4}), (\ref{4.5}). We find
\be \begin{array}{c}
     \d\xi^i \d \xi^j + \d\xi^j \d\xi^i = \d t \d b^{ij}\;, \\
     \d t \d t =0\;, \quad\qquad\d\xi^i \d t + \d t \d\xi^i=0\;.
 \end{array}\label{4.14}
\ee
These are special cases of
\be \d f\d g+\d g\d f= \d t\,\d b(\d f,\d g)\;, \label{4.15} \ee
which is obtained by application of $\d$ on (\ref{4.6}).

{\em From (\ref{4.6}) and (\ref{4.15}) follows that all deviations of the
present differential calculus from the classical differential calculus
are proportional to $\d t$. Therefore and by the second equation in
(\ref{4.14}) it is also clear that for forms $\d t\phi$, for any $\phi\in\O$
all calculations proceed classically.} This will help us to proceed more
rapidly in what follows.

We extend the $\bu$ product to act between any 1-form $\alpha$ and an
arbitrary form $\phi$ by  using the ``insert'' opeartor $\ins$ of ordinary
exterior calculus
\be   \alpha\bu\phi:=\d t\,\alpha^b\ins\phi\;. \label{4.16} \ee
On the right hand side everything is as in the ordinary differential calculus
because of the presence of $\d t$ . It is not difficult to see that
(cf. (\ref{4.2}))
\bez   \d f\bu\phi =[\phi,f]\;. \eez
For a one form $\alpha$ the combination $\alpha\bu$ acts as a derivation
of the product of differential forms, i.e.
\bez \alpha\bu(\phi\psi)=(\alpha\bu\phi)\psi+\phi(\alpha\bu\psi)\;.\eez
The elements $u$ of $\X$ which vanish on $t$, i.e.\ $u(t)=0$ are derivations
of $\A$ and define a left $\A$ submodule $\X_1$ of $\X$. With every
$u\in\X_1$ we associate mappings $D_u:\O\to\O$ defined up to terms
lying in $\d t\,\O$  through the following relations
\be  D_u(\phi\psi)=(D_u\phi)\psi+\phi(D_u\psi)\pmod{\d t}\;,\label{4.17}\ee
\be  D_u\,f:=u(f)\;,\label{4.18}\ee
and
\be D_u\d t=0\pmod{\d t}\;. \label{4.19}\ee
We write $D_i$ for $D_{\pa_i}$ and we set
\be D_i\d\xi^j =-\d\xi^k\,\Gamma^j{}_{ki}\pmod{\d t}\;, \label{4.20}\ee
for the coefficients of the connection. For an $r$-form $\phi$
with
\bez
\phi={1\over r!}\phi_{i_1\cdots i_r}\d\xi^{i_1}\cdots\d\xi^{i_r}
  \pmod{\d t}\;,
\eez
we find
\be
 D_k\phi={1\over r!}\na_k\phi_{i_1\cdots i_r}\d\xi^{i_1}\cdots\d\xi^{i_r}
  \pmod{\d t}\;, \label{4.21}
\ee
where
\be
    \na_k\phi_{i_1\cdots i_r}:=\pa_k\phi_{i_1\cdots i_r}-
     \Gamma^j{}_{ki_1}\phi_{j\cdots i_r}-\cdots-
     \Gamma^j{}_{ki_r}\phi_{i_1\cdots j}\;.\label{4.22}
\ee
Extending these definitions as usual to tensor products we obtain
\be
\na_i b^{jk}:=\pa_i b^{jk}+b^{\ell(j}\Gamma^{k)}{}_{\ell i};. \label{4.23}
\ee
In the following {\em we demand the connection to be torsion free i.e.}
$\Gamma^i{}_{[jk]}=0$, from which follows
\be     \d\phi =\d\xi^i\,D_i\phi\pmod{\d t}\;, \label{4.24} \ee
and {\em $b$-compatible}, that is $\na_ib^{jk}=0$.

It should be emphasized here that in the context of the present
noncommutative differential calculus it is possible to develop systematically
tensor analysis so that the introduction of the above mentioned concepts
of a connection and covariant derivative (cf.\  eqs(\ref{4.17})--(\ref{4.20}),
(\ref{4.21})--(\ref{4.23})) is made perfectly rigorous. However, this would
lead us far away from our task to develop symplectic geometry and Hamiltonian
dynamics and therefore it will be presented in a subsequent paper.
Let us also remark here that in the old-fashioned index notation these
differential
geometric tools are first introduced in \cite{40}.

With the aid of these mappings and the $\bu$ we define a new product in
$\O$
\be   \phi\we\psi:=\phi\psi+(D_i\phi)(\d\xi^i\bu\psi)\;.\label{4.25} \ee
It is easy to check that $\we$ is right $\A$ linear in both factors, i.e.
\be  (\phi f)\we(\psi g)=(\phi\we\psi)fg\;.\label{4.26} \ee
Furthermore it can be shown that the product is associative and as we shall
see below also graded commutative, i.e.\ for $\phi\in\O^r$
and $\psi\in\O^s$
\bez    \phi\we\psi=(-1)^{rs}\psi\we\phi\;. \eez
An easy consequence of the above definition is
\bez
 \phi\we f=\phi f\;,\qquad\qquad
 f\we\phi=f\phi+\d f\bu\phi=\phi f\;.
\eez
We define now an operator
\be \D f:=\d f+{1\over 2}\d\xi^i\bu D_i\d f\;,\label{4.27} \ee
motivated by the fact that it satisfies the usual Leibniz rule
\bez   \D(fg)=(\D f)g+(\D g)f\;\eez
Note that because of this property the 1-forms
\be     \D\xi^i:=\d\xi^i-{1\over2}\d t \Gamma^i\;,\label{4.28} \ee
transform right-covariantly under a change of coordinates $\xi'{}^j
=\xi'{}^j(\xi)$, i.e.\ we have $\D'\xi'{}^j=\D\xi^i\,(\pa_i\xi'{}^j)$.
Here we have set $\Gamma^i:=b^{jk}\Gamma^i{}_{jk}$. Clearly $\d t,\,\D\xi^1,
\ldots,\D\xi^{2n}$ form a basis of $\O^1$ with
\bez  \D f=\D\xi^i(\pa_i f)+\d t(\pa_t f)\;,\eez
and
\be \d f & = & \D f - {1\over2}\d t \,b^{ij}\na_i\pa_j f\;,\nonumber\\
        & = & \D\xi^i(\pa_i f)+\d\xi^i (\tp_t f)\;, \label{4.29}\ee
where
\be   \tp_t f:=\pa_t f -{1\over 2} b^{ij}\na_i\pa_j f\;.\label{4.30}\ee
The vector fields $\tp_t,\,\pa_1,\ldots,\pa_{2n}$ form a basis of $\X$ dual
to the above basis of $\O^1$. From the definitions we  find
\be
 \D\xi^i\D\xi^j=\d\xi^i\d\xi^j+{1\over2}\d t \d\xi^{[i}\Gamma^{j]}\;,
\label{4.31}\ee
\be
\D\xi^i\we\D\xi^j=\D\xi^i\D\xi^j+\d t\d\xi^k\,
    b^{\ell j}\Gamma^i{}_{k\ell}\;,\label{4.32}
\ee
and using (\ref{4.14}) we obtain
\be \begin{array}{c}
   \D\xi^i\we\D\xi^j+\D\xi^j\we\D\xi^i=0\;,\\
 \d t\we\D\xi^i=\d t\d\xi^i\;,\qquad \D\xi^i\we\d t=\d\xi^i\d t\;,
\qquad \d t\we \d t =\d t\d t=0\;.\end{array}\label{4.33}
\ee
It is now easy to see using (\ref{4.26}) and (\ref{4.33}), that $\we$ is
antisymmetric. With the aid of the curvature of the connection $\Gamma$,
\bez
    R^i{}_{jk\ell}:=\pa_k\Gamma^i{}_{j\ell}-\pa_\ell\Gamma^i{}_{jk}
     +\Gamma^i{}_{mk}\Gamma^m{}_{j\ell}-\Gamma^i{}_{m\ell}\Gamma^m{}_{jk}\;,
\eez
and the curvature 2-form
\bez \O^i{}_j:={1\over2} R^i{}_{jk\ell}\d\xi^k\d\xi^\ell\pmod{\d t}\;, \eez
we can prove the following usefull formulae
\bez
 D_{[i}D_{j]}\phi=-R^k{}_{\ell ij}\d\xi^\ell(\pa_k\ins\phi)\pmod{\d t}\;,
\eez
\bez
 \d D_i\phi-D_i\d\phi=-\O^j{}_i(\pa_j\ins\phi)+
    \d\xi^j\Gamma^k{}_{ij}D_k\phi\pmod{\d t}\;,
\eez
and
\bez
  \d(\d\xi^i\bu\phi)-\d\xi^i\bu\d\phi=-\d t\,b^{ij}D_j\phi-
    \d\xi^j\Gamma^i{}_{jk}(\d\xi^k\bu\phi)\;.
\eez
Using these one can prove (see also \cite{28} \S7) for $\phi\in\O^r$ and
$\psi\in\O$
\be
  \d(\phi\we\psi)=(\d\phi)\we\psi+(-1)^r\phi\we(\d\psi)+
    \O^j{}_i(\pa_j\ins\phi)(\d\xi^i\bu\psi)+
    \d t\,b^{ij}(D_i\phi)(D_j\psi)\;.\label{4.34}
\ee
A special case of this formula is
\be
  \d(\phi f)=(\d\phi)f+(-1)^r\phi\we\d f+\d t\,b^{ij}(D_i\phi)(\pa_j f)\;.
\label{4.35}\ee
For a 1-form $\alpha=\D\xi^i\alpha_i+\d t \alpha_t$ we find using (\ref{4.34})
\bez
  \d\alpha={1\over2}\D\xi^i\we\D\xi^j\,\pa_{[i}\alpha_{j]}+\d t\d\xi^i
  \left[\pa_t\alpha_i-\pa_i\alpha_t-{1\over2}b^{jk}(\na_j\na_k\alpha_i
   +R^\ell{}_{jki}\alpha_\ell)\right]\;.
\eez
To connect the above results to symplectic geometry and Hamiltonian dynamics
we need the exterior derivative of a 2-form $\omega$. In general
\bez
  \omega={1\over2}\D\xi^i\we\D\xi^j\,\omega_{ij}+\d t\d\xi^i\omega_i\;,
\eez
(see (\ref{4.33}) and the remark before (\ref{4.29})). Then a lengthy
calculation gives
\be
  \d\omega &=& {1\over3!}\D\xi^i\we\D\xi^j\we\D\xi^k\,\left[{1\over2}\pa_{[i}
  \omega_{ik]}\right]+ \nonumber\\
&&\!\!{1\over2}\d t\d\xi^i\d\xi^j\left[\pa_t\omega_{ij}-
  \pa_{[i}\omega_{j]}-{1\over2}b^{k\ell}(\na_k\na_\ell\omega_{ij}-
 R^m{}_{k\ell[i}\omega_{j]m}-R^m{}_{kij}\omega_{\ell m})\right]\:.
\label{4.36}
\ee
Therefore a 2-form $\omega$ is closed, if
\be    \pa_{[i}\omega_{jk]}=0\;, \label{4.37} \ee
and
\be
  \pa_t\omega_{ij}-\pa_{[i}\omega_{j]}-{1\over2}b^{k\ell}
  (\na_k\na_\ell\omega_{ij}-R^m{}_{k\ell[i}\omega_{j]m}-
  R^m{}_{kij}\omega_{\ell m})=0\;.\label{4.38}
\ee
If we now assume as in ordinary symplectic mechanics, that
$\pa_t\omega_{ij}=0$ then it can be proved that (\ref{4.37}), (\ref{4.38})
imply that $\omega_i$ satisfies
\be
   \pa_{[i}(\omega_{j]}+{1\over2}b^{k\ell}\na_{|k}\omega_{\ell|j]})=0\;,
\label{4.39}\ee
where only $i,j$ are antisymmetrized. Therefore $\omega$ takes the form
\bez
  \omega={1\over2}\D\xi^i\we\D\xi^j\,\omega_{ij}
   +\d t\d\xi^i(\pa_i H-{1\over2}b^{k\ell}\na_{ k}\omega_{\ell i})\;,
\eez
for some function $H$. On the other hand (\ref{4.37}) is the condition for
$\omega_{ij}$ to be closed in the ordinary exterior calculus, hence by
Darboux's theorem it can be brought locally to the form
\bez
  \omega_{ij}=J_{ij}\;,\qquad
  J=(J_{ij})=\left(\begin{array}{cc} 0 & I \\  -I & 0 \end{array}\right)\;,
  \qquad I=(\delta_{\alpha\beta})\;,\;\; \alpha,\beta=1,\ldots,n\;,
\eez
where $J$ is the symplectic form in canonical coordinates.  According to the
interpretation of $\xi^i$ in section 5 this can also be assumed in the present
context. Therefore, to give our results a more familiar form we write
$\omega$ in this form
\be
  \omega={1\over2}\D\xi^i\we\D\xi^j\,J_{ij}
   +\d t\d\xi^i(\pa_i H-{1\over2}b^{k\ell}\na_{ k}J_{\ell i})\;.\label{4.40}
\ee
However the subsequent calculations do not make any use of the fact that
$\xi^i$ are taken here to be canonical coordinates.

We already used the ``insert'' operator $\ins$ in the sense of the classical
differential calculus. It is necessary to extend its definition in the present
context, since it is needed in the calculation of Hamiltonian vector fields.
For $X\in\X$, $f\in\A$ it is natural to put $X\ins f=0$ and as in the
ordinary exterior calculus we set
\be \begin{array}{c}
     X\ins\alpha := \< X,\alpha\>\;,\\
  X\ins(\phi\we\psi) :=	(X\ins\phi)\we\psi+(-1)^r\phi\we(X\ins\psi)\;,
\end{array}\label{4.41}
\ee
for $\alpha\in\O^1$, $\phi\in\O^r$ and $\psi\in\O$.
If $X=X^i\pa_i+X^t\tp_t$ in the coordinate system in which $\omega$ has the
form (\ref{4.40}), we have $X\ins\D\xi^i=X^i$ and $X\ins\d t=X^t$.
Therefore we find that
\be
  X\ins\omega=\D\xi^i\left[-J_{ij}X^j+X^t(\pa_i H+F_i)\right]+
   \d t X^i(\pa_i H +F_i)\;,\label{4.42}
\ee
\be F_i:=-(1/2)b^{jk}\na_j J_{ki}\;. \label{4.43}\ee
In order to write $\omega$ in the form (\ref{4.40}), it is necessary
that it has maximal rank, which is $2n$ since $M\times\Rl$ is
odd-dimensional. Therefore it has a 1-dimensional kernel given by the relation
\be X\ins\omega=0\;.\label{4.44} \ee
In ordinary extended Hamiltonian mechanics $H$ is the
Hamiltonian\footnote{Notice that by (\ref{4.31}), (\ref{4.32}),
eq.(\ref{4.40}) is $\omega=(1/2)\d\xi^i\d\xi^j J_{ij}+\d t \d\xi^i\pa_i H$,
strongly reminding conventional extended Hamiltonian dynamics.} and
the kernel is identified by definition with the space of Hamiltonian
vector fields $X$.

Equations (\ref{4.42}), (\ref{4.44}) give
\be
J_{ij} X^j=X^t(\pa_i H+F_i)\;,\qquad\qquad X^i(\pa_i H+F_i)=0\;. \label{4.45}
\ee
Setting $J^{ij}:=J_{ij}$ we have $J^{ik}J_{jk}=\delta^i_j$ hence the first
equation gives
\bez   X^i=-X^t J^{ij}(\pa_jH+F_j)\;, \eez
and consequently  the 2nd equation is identically satisfied. Therefore the
Hamiltonian vector field defined by $H$ is given by
\be
 X=X^t\left( (\pa_iH+F_i)J^{ij}\pa_j+(\pa_t-{1\over2}b^{ij}\na_i\pa_j)\right)
\;. \label{4.46}
\ee
As in ordinary extended Hamiltonian dynamics, the equation of motion for an
observable $A$, i.e.\ for $A\in\A$ takes the form $X\,A=0$, which gives
\be
  \pa_t A=- \left[ \{H,A\}+F_i J^{ij}\pa_j A \right]+{1\over2}
   \pa_i(b^{ij}\pa_j A)+{1\over2}b^{ij}\Gamma^k{}_{ki}\pa_j A\;.\label{4.47}
\ee
The {\em Hamiltonian equation} is now identical with the general kinetic
equation (\ref{2.5}) for a classical open system in interaction with a
large bath at canonical equilibrium, provided that
\newcounter{saveeqn}
\setcounter{saveeqn}{\value{equation}}
\stepcounter{saveeqn}
\setcounter{equation}{0}
\renewcommand{\theequation} {\arabic{section}.\arabic{saveeqn}\alph{equation}}
\be
    \alpha^{ij} & = & {1\over2}b^{ij}\;, \label{4.48a}\\
    F_i & = & \pa_i F\;, \label{4.48b} \\
    \Gamma^k{}_{ki} & = & -\beta\pa_i H\;, \label{4.48c}
\ee
\setcounter{equation}{\value{saveeqn}}
\renewcommand{\theequation} {\arabic{section}.\arabic{equation}}
with $\{F,H\}=0$.

Condition (\ref{4.48b}) is equivalent to $\pa_{[i}F_{j]}=0$ which by
(\ref{4.38}), (\ref{4.39}), (\ref{4.43}) takes the form
\be
 b^{k\ell}(\na_k\na_\ell J_{ij}-R^m{}_{k\ell[i}J_{j]m}-
        R^m{}_{kij}J_{\ell m})=0\;.\label{4.49}
\ee
When $b^{ij}$ is nondegenerate, this is equivalent to the condition that the
{\em symplectic form $J_{ij}$ is harmonic with respect to the Laplace-Beltrami
operator of $b$ } (see e.g.\ \cite{34} p.3). Equation (\ref{4.48c}) is
equivalent to the condition that the canonical measure
\be\epsilon=e^{-\beta H}\d\xi^1\cdots\d\xi^{2n}\pmod{\d t}\;,\label{4.50}\ee
is covariantly constant $D_i\epsilon=0$ (see e.g.\ \cite{35} p.215).

The above results can now be summarized by saying that if the phase-space
of a classical open system $\Sigma$ is endowed with a noncommutative
geometrical structure (\ref{4.4}), (\ref{4.5}), because of its interaction
with a bath at canonical equilibrium, then the corresponding {\em Hamiltonian}
evolution of observables is identical to that given by conventional kinetic
theory {\em if} the symplectic form is ``harmonic'' with respect to the
``metric'' connection defined by $b^{ij}$, and the canonical
(Maxwell-Boltzmann) measure defined on $\Sigma$ at the bath temperature is
covariantly constant. Condition (\ref{2.3}) is not expected to follow from
the procedure followed so far, unless a precise relation of (\ref{4.4}),
(\ref{4.5}) to conventional dynamics is somehow made plausible. This will
be considered in another paper.

\section{Discussion}
\setcounter{equation}{0}

In the previous section we have given geometric conditions so that Hamiltonian
dynamics in the context of noncommutative differential calculus defined on
$M\times\Rl$ by (\ref{4.4})-(\ref{4.6}) can be interpreted physically as
kinetic theory of classical open systems interacting with a large bath at
canonical equilibrium.

As already remarked at the end of section 3 the presentation so far is
formal in the sense that the nature of the algebra $\A$ and its corresponding
coordinate representation in terms of $\xi^i$ has not been specified. Here we
discuss these questions further, but it should be emphasized that this is not
done rigorously. Actually much remains to be done for the complete
clarification of the problems addressed in this section.

At the beginning of section 4 we remarked that a 1-form $\alpha$ in the
universal differential envelope of the algebra of functions on a set $N$ is
a function $\alpha:N\times N\to\Cx$ obtained by the obvious extension of
(\ref{4.1}), that is of
\be   (f\~{\d}g\,h)(a,b)=f(a)[g(b)-g(a)] h(b)\;.\label{5.1}\ee
For the $\~{\bu}$ we can show that (\ref{5.1}) implies
\be
 (\~{\d}f_1\~{\bu}\cdots\~{\bu}\~{\d}f_r)(a,b)=
 (f_1(b)-f_1(a))\cdots(f_r(b)-f_r(a))\;.\label{5.2}
\ee
Relations like $\alpha\~{\bu}\beta-\~{\d}t\,\~{b}(\alpha,\beta)=0$
are in general incompatible with the above prescription of evaluating
differential forms. Therefore if we do impose such relations (i.e.\ pass
from $(\~{\O},\~{\d},\~{\bu})$ to $(\O,\d,\bu)$ as it is outlined at the
beginning of section 4) {\em and} at the same time we still want to
retain somehow an interpretation of $\bu$, similar to that given by
(\ref{5.2}) then the elements of $\O^1$ cannot be functions on the whole of
$N\times N$. In fact such relations induce some structure on $N\times N$ by
grouping together points of $N$ which may be considered as neighbouring.
This is best illustrated by giving some examples.

Take $N:=\Rl$ the real line. Let $x$ be the coordinate function on $N$ and
impose the relation $\d x\bu\d x -\d x=0$ (cf.\ \cite{41}). If we want to
keep the interpretation of one forms as functions on some set $N_1$, this
cannot be the whole of $N\times N$. It is obvious that $N_1$ must be that
subset of $N\times N$ on which the imposed condition is satisfied identically.
If $(a,b)\in N_1 \subset N\times N$ then since $x(a)=a$ we find
\bez
    0=(\d x\bu\d x-\d x)(a,b)=(b-a)(b-a-1)\;.
\eez
Hence in order for $(a,b)$ to be an element of $N_1$ either $b=a$ or
$b=a+1$. Hence $N_1=\{(a,a),(a,a+1)|\,a\in\Rl\}$. This set gives a structure
on the set $N$ by specifying which of its points are to be considered as
neighbouring. Obviously the above condition specifies a discrete structure
on $\Rl$. The possibility to evaluate a one form on $(a,b)$ with $a$ and
$b$ not neighbours is still given, if $b-a=m\in\Ir$, and corresponds to the
``integral''
\bez  \int^b_a\alpha:=\sum_{k=1}^m \alpha(a+k-1,a+k)\;.\eez
Applying the same reasoning to the relation $\d f\bu\d g=0$ with
smooth functions $f,g:\Rl\to\Cx$, then for $a,b\in\Rl$ we find
\bez
 (\d f\bu\d g)(a,b)=\left(f(b)-f(a)\right)\left(g(b)-g(a)\right)=
 f'(x_1)g'(x_2)(b-a)^2=0\;,
\eez
where we have used the mean value theorem. Since this condition must hold
for all smooth functions we must have $\epsilon:=b-a$ and $\epsilon^2=0$.
Now this relation gives something trivial since $\epsilon$ must identically
vanish. But the relation $\d f\bu\d g=[\d f,g]=0$ holds in the usual
differential
calculus and consequently it cannot be trivial. In fact one may interprete
relation $\epsilon^2=0$ as saying, that $\epsilon$ is an infinitesimal
of first order. In this sense $\Rl$ is again structured since now
$N_1:=\{(a,a),(a,a+\epsilon)|\,a\in\Rl\}$. For arbitrary $a<b\in\Rl$ the
integral is defined by
\be \int^b_a\alpha := \lim\sum_{k=1}^m\alpha(x_{k-1},x_k)\;,\label{5.3}\ee
where this is obtained by taking the limit of vanishing width of the partition
$a=x_0<x_1<\cdots<x_m=b$ of $[a,b]$. Now the relation $\d f\bu\d g=0$
integrated over $[a,b]$ for arbitrary $a<b\in\Rl$ gives
\be
 0=\int^b_a \d f\bu\d g=\lim\sum_{k=1}^m
   \left(f(x_k)-f(x_{k-1})\right)\left(g(x_k)-g(x_{k-1})\right)\;,\label{5.4}
\ee
which can be expressed by saying that the ``quadratic variation''
of functions must vanish. This is true {\em if} $f,g$ are of bounded variation,
a condition which is {\em necessary} for the Riemann-Stieltjes integral
$\int^b_a g\d f$ to exist.

For a second order calculus on smooth functions of one variable $\xi$,
parametrizing $N:=\Rl$, we have by definition (section 4, eq.(\ref{4.7}))
that $\d f\bu\d g\bu\d h=0$, hence by applying $\d\xi\bu\d\xi\bu\d\xi=0$ on
$(a,b)\in\Rl^2$ we get $(\xi(b)-\xi(a))^3=0$. Thus $\xi(b)-\xi(a)$ is an
infinitesimal of {\em second} order. Therefore  $N_1=\{(a,a),(a,a+\epsilon),
(a,a+\epsilon^2)|\,a\in\xi^{-1}(\Rl)\}$, in this case and consequently for
given $a\in N$ we can move away from $a$ in two ways, either to $a+\epsilon$
or to $a+\epsilon^2$. In this sense then $N$ becomes structured and can be
considered as 2-dimensional.\footnote{The equality $N=\Rl$ is only
set-theoretic.}

For $a<b\in\Rl$ we define formally an integral as in (\ref{5.2}). Applying
this on $\d\xi\bu\d\xi\bu\d\xi=0$ we obtain
\be \lim \sum_{k=1}^m\left(\xi(x_k)-\xi(x_{k-1})\right)^3=0\;,\label{5.5}\ee
where again the limit is obtained as in (\ref{5.3}). This relation can be
expressed by saying that the ``cubic variation'' of $\xi$ must vanish. It
is perhaps of independent mathematical interest to find equivalent
characterizations of such functions.

Since the quadratic variation of $\xi$ does not vanish in general, $\xi$
cannot be the usual coordinate function of $\Rl$. Consequently we have $\Rl$
as a differentiable manifold but with a differential structure which is
not the standard one. If we set $\d t:=(1/b)\d\xi\bu\d\xi$ with some constant
$b$ and $t$ a function on $\Rl$, then by (\ref{2.5}) $\d t\bu\d t=0$ and
hence $t$ is of bounded variation and can be taken to be the coordinate
function on some copy of $\Rl$. This additional coordinate $t$ realizes
somehow the fact that $N$ is 2-dimensional.

In the light of the above remark, if $M=\Rl^{2n}$, it seems that $\xi^i$
should be interpreted as local coordinates defined on $\Rl^{2n}$ by an atlas
$\hat{\cal U}$ not compatible with the usual one giving the standard
differential structure of $\Rl^{2n}$; that is there are charts in
$\hat{\cal U}$  not $C^k$-related to the identity mapping of $\Rl^{2n}$,
for some $k>0$. Equivalently  we may say that the identity mapping of
$\Rl^{2n}$, does not belong to $\hat{\cal U}$.
More generally, for a $2n$-dimensional differentiable manifold $M$ the above
discussion implies that if $\xi$ is a local chart of $M$ in $\Rl^{2n}$ and
$\hat{\xi}$ a local chart of $\Rl^{2n}$ from $\hat{\cal U}$ the functions
$f:M\to\Cx$ belonging to $\A$, are smooth functions of $\xi$ but are not
smooth, not even of bounded variation, as functions of $\hat{\xi}$,
that is $f\circ\xi^{-1}$ are smooth, but $f\circ(\hat{\xi}\circ\xi)^{-1}$
are not.

The whole discussion in this section reminds us strongly of stochastic
calculus on manifolds developped in the context of semimartingale theory,
in particular when stochastic terms are given in terms of Wiener processes
(see e.g.\ \cite{36}, \cite{37}). In fact there are many results in
stochastic calculus having an exact analogue in the formalism developped
here. As examples compare (\ref{4.8}) with (4) in \cite{37} p.134 and
their properties, or elements of $\X$ eq.(\ref{4.10}) with the
characterization of 2nd order fields in \cite{37} Lemma 6.1. Moreover
our relation of the ``metric connection'' with the drift term in the general
Fokker-Planck type kinetic equation (\ref{4.43}) suggests a close relation
with the interpretation in stochastic calculus of a connection on a manifold
as a mapping giving the ``drift'' of a 2nd order vector field
(\cite{36} p.258--259).In fact it seems possible --- and it will be examined
elsewhere --- that our present model of noncommutative geometry can be
realized in the context of stochastic calclulus. However, whether this is the
only possibility remains an interesting, but to our knowledge, still unsolved
problem.

\vskip.5cm
\noindent
{\bf Acknowledgment.} We would like to thank D.~Ellinas and
F.~M\"uller-Hoissen for stimulating discussions and  critical
comments, and H.~Goenner for his interest in this work.


\small
\begin{thebibliography}{99}
\bibitem{1}  van Kampen N G  1981 {\em Stochastic processes in physics and
  chemistry}, North-Holland.
\bibitem{2} Chandrasekhar S  1943 Rev.\ Mod.\ Physics {\bf 15} 1
\bibitem{2a} Wax N 1954 {\em Selected papers on noise and stochastic
  processes} Dover pp.93, 319
\bibitem{3} Gardiner C W 1985 {\em  Handbook of stochastic methods} Springer
\bibitem{4} Arnold L 1973 {\em Stochastic differential equations} Wiley
\bibitem{5} Balescu R 1963 {\em Statistical mechanics of charged particles}
  Wiley
\bibitem{6} Balescu R 1975 {\em Equilibrium and nonequilibrium statistical
  mechanics} Wiley
\bibitem{7} R\'{e}sibois P and de Leener M 1977 {\em Classical kinetic theory
  of fluids} Wiley
\bibitem{8} Ichimaru S 1991 {\em Statistical plasma physics} Vol.I
  Addison-Wesley
\bibitem{9} Liboff R L 1969 {\em Introduction to the theory of kinetic
  equations}  Wiley
\bibitem{10} Zwanzig R W 1960 in {\em Lectures in Theoretical physics} Vol.3
  Summer Institute of Theoretical Physics, University of Colorado
\bibitem{11} Tzanakis C and Grecos A P 1988 Physica {\bf A149} 232
\bibitem{12} Grecos A P and Tzanakis C 1988 Physica {\bf A151} 61
\bibitem{13} Bogoliubov N N 1977 {\em On the stochastic processes in the
  dynamical systems} Joint Institute of Nuclear Researsch Dubna eq (2.27)\\
   Grigolini P 1988 in {\em Noise in non-linear dynamical systems} Vol.I
   ch.5  eq.(58)  Mossand F and McClintoc P V E (eds)\\
   Severne G 1965 Physica {\bf 31} 877 eq.(4.2.3)\\
   Haggerty M J and Severne G 1976 Adv.\ Chem.\ Physics {\bf 35} 119
    eqs(3.34) and (3.24), (3.35)\\
   Balescu R and Misguich J H 1974 J.\ Plasma Physics {\bf 11}  377
   eq.(4.10)\\
   Balescu R and Misguich J H 1975 J.\ Plasma Physics {\bf 13} 385
   eq.(4.2.2) together with (4.1.5), (4.1.16), (4.1.18)\\
   Lindenberg K and Mohanty V 1983 Physica {\bf A119} 1  eq.(A.20)\\
   van Kampen N G 1976 Phys.\ Reports {\bf 24} No 3 eq.(10.4)
\bibitem{14} Fedyanin V K and Gavrilenko G M 1979 Physica {\bf A99} 34
   eqs (35), (36)\\
   Kandrup H E 1981 Ap.\ J.\ {\bf 244} 316  eqs (37)-(39)\\
   Leaf B 1972 Physica {\bf 58} 445
\bibitem{15} Davies E B 1976 {\em Quantum theory of open systems} Academic
  Press
\bibitem{16} Davies E B 1974 Comm.\ Math.\ Physics {\bf 39} 91\\
   Davies E B 1976 Math.\ Annalen {\bf 219} 147
\bibitem{17} Spohn H and Lebowitz J L 1978 Adv.\ Chem.\ Physics {\bf 38} 109
\bibitem{18} Lindblad G 1976 Comm.\ Math.\ Physics {\bf 48} 119
\bibitem{19} D\"umcke R and Spohn H 1979 Z.\ Physik {\bf B34} 419
\bibitem{20} Reed M and Simon B 1974 {\em Methods of modern mathematical
   physics} Academic Press Vol.I
\bibitem{21} Davies E B 1980 {\em One-parameter semigroups} Academic Press
\bibitem{22} Yosida K  1971 {\em Functional analysis} Springer
\bibitem{23} van Casteren J A 1985 {\em Generators of strongly continuous
  semigroups} Pitnam Publishing Inc.
\bibitem{24} Nelson E 1967 {\em Dynamical theories of Brownian motion}
  Princeton
\bibitem{25} Reed M and Simon B (1978) {\em Methods of modern mathematical
  physics} Academic Press Vol.IV
\bibitem{26} Tzanakis C 1995 ``Classical kinetic equations: Explicit form and
  its mathematical and physical foundations'' Preprint
\bibitem{27} Tzanakis C 1988 Physica {\bf A151} 90
\bibitem{28} Dimakis A and M\"uller-Hoissen F 1993 Lett.\ Math.\ Physics
  {\bf 28} 123
\bibitem{29} Baehr H, Dimakis A and M\"uller-Hoissen F 1995 J.\ Phys.\ A
  {\bf 28} 3197
\bibitem{30} Dimakis A and M\"uller-Hoissen F 1994 in the Proceeding of the
  XXII International Conference on Differential Geometric Methods in
  Theoretical Physics, Keller J and Oziewics Z (eds), Universidad Nacional
  Autonoma de M\'exico
\bibitem{31} Frigerio A and Gorini V 1984 J.\ Math.\ Physics {\bf 25} 1050
\bibitem{32} Spohn H 1980 Rev.\ Mod.\ Physics {\bf 53} 569
\bibitem{33} Pawula R F 1967 Phys.\ Rev.\ {\bf 162} 186
\bibitem{34} Lichnerowicz A 1977 {\em Geometry of groups of transformation}
  Noordhoff International Publishing
\bibitem{35} Schutz B 1980 {\em  Geometrical methods of mathematical physics}
  Cambridge U.P.
\bibitem{36} Meyer P A 1981 in {\em Stochastic Integrals} Williams D (ed.)
  Lecture Notes in Mathematics Vol.851 Springer
\bibitem{37} Emery M 1989 {\em Stochastic calculus in manifolds} Springer
\bibitem{38} Bismut J-M 1981 {\em M\'ecanique Al\'eatoire} Lecture Notes in
  Mathematics Vol.866 Springer
\bibitem{39} Connes A and Lott J 1990 Nucl. Physics B (Proc.\ Supp.) {\bf 18}
  29
\bibitem{40} Dimakis A and M\"uller-Hoissen F 1993 Int.\ J.\ Mod.\ Physics A
  (Proc. Suppl.) {\bf 3A} 474; 1992 ``Noncommutative differential
  calculus, gauge theory and gravitation'' Preprint G\"ottingen GOE--TP
  33/92
\bibitem{41}  Dimakis A, M\"uller-Hoissen F and Striker T 1993 Phys.\ Lett.\
   {\bf 300B} 141; 1993 J.\ Phys.\ A {\bf 26} 1927
\end{thebibliography}


\normalsize
\end{document}